\documentclass[12pt]{article}

\usepackage{amsfonts,amssymb,chet,dsfont,mathrsfs,pgfplots,tikz,graphicx}

\def\beqn{\begin{eqnarray}} 
\def\eeqn{\end{eqnarray}} 
\def\be{\begin{equation}}
\def\ee{\end{equation}}
\def\nn{\nonumber}

\def\nn{\nonumber}

\usetikzlibrary{decorations.markings,arrows,calc}
\tikzset{>=latex,->/.style={decoration={markings,mark=at position 1 with {\arrow[scale=1.5]{>}}},postaction={decorate}}}
\tikzset{->-/.style={decoration={markings,mark=at position 0.6 with {\arrow[scale=1.5]{>}}},postaction={decorate}}}
\tikzset{-<-/.style={decoration={markings,mark=at position 0.6 with {\arrow[scale=1.5]{<}}},postaction={decorate}}}

\preprint{BONN-TH-2013-18\\
CERN-PH-TH/2013-242\\
SU-ITP-13/19\\
CETUP2013-017\vspace{-1.6cm}
}

\title{Supernova Constraints on MeV Dark Sectors\\
from $e^+ e^-$ Annihilations}

\author{Herbert K. Dreiner$^{\ast,}$\email{dreiner@th.physik.uni-bonn.de},
Jean-Fran\c{c}ois Fortin$^{\dagger,\$,}$\email{jean-francois.fortin@cern.ch},
Christoph Hanhart$^{\S,}$\email{c.hanhart@fz-juelich.de}\\
and Lorenzo Ubaldi$^{\ast,}$\email{ubaldi@th.physik.uni-bonn.de}
} 

\affiliation{$^\ast$Physikalisches Institut der Universit\"{a}t Bonn, Nussallee 12, D-53115 Bonn, Germany\\
$^\dagger$Theory Division, Department of Physics, CERN, CH-1211 Geneva 23, Switzerland\\
$^\$$Stanford Institute for Theoretical Physics, Department of Physics, Stanford University, Stanford, CA 94305, USA\\
$^\S$Institut f\"{u}r Kernphysik, Forschungszentrum J\"{u}lich, 52428 J\"{u}lich, Germany
}

\abstract{Theories with dark forces and dark sectors are of interest for dark matter models.  In this paper we find the region in parameter space that is constrained by supernova cooling constraints when the models include dark sector particles with masses around $100$ MeV or less.  We include only interactions with electrons and positrons.  The constraint is important for small mixing parameters.
}

\date{October 2013} 

\begin{document}

\maketitle


\section{Introduction}\label{Intro}

Theories with dark forces~\cite{Fayet:1990wx, Holdom:1985ag} are well-motivated extensions of the Standard Model (SM).  Such extensions might provide an explanation for dark matter (DM), if it is assumed that new stable particles charged under the dark gauge group exist. Gauge kinetic mixing then generates interactions between the dark sector particles and SM particles.  To constrain such very weakly-coupled models with light dark sector particles, it is convenient to study the dark sector particle production mechanisms in astrophysical bodies such as white dwarfs (WDs) and supernovae (SN).  Recently, some of us studied these constraints in WDs and determined that interesting parts of the parameter space which are or will be probed by experiments are already mostly excluded when the dark sector particle masses are $\sim\mathscr{O}(\text{few tens of keV})$~\cite{Dreiner:2013tja}.  Note that such light particles might already be problematic for big bang nucleosynthesis (BBN).  However, BBN constraints suffer from several caveats that do not apply to WD constraints~\cite{Steigman:2013yua}.  More importantly though, dark sector particles with masses in the MeV range, inaccessible to WDs, are more interesting because they could provide a viable DM candidate~\cite{Pospelov:2007mp} and explain the 511 keV line from the galactic center observed by INTEGRAL~\cite{Knodlseder:2003sv}. 
 Since temperatures reach $\sim\mathscr{O}(\text{few tens of MeV})$ inside SN, it is thus natural to investigate SN constraints on such theories.

The idea behind the astrophysical bounds on new particles is simple: if new particles are light enough to be produced in astrophysical bodies, they can possibly escape and generate excess cooling.  This could contradict the agreement between theoretical cooling models and observations.  Since SN contain electron-positron pairs ($e^-/e^+$) as well as nucleons ($N$), the possible dark sector particle (Dirac fermion $\psi$ and/or complex scalar $\phi$) production mechanisms are
\eqna{e^++e^-&\to\genfrac{\{}{.}{0pt}{}{\bar{\psi}+\psi}{\phi^\dagger+\phi}\\
N+N&\to\genfrac{\{}{.}{0pt}{}{N+N+\bar{\psi}+\psi}{N+N+\phi^\dagger+\phi.}
}
Once produced, the dark sector particles escape the SN if their mean free path $\lambda_{\psi,\phi}$ is large enough, of the order of the SN core.  The scattering processes of interest in SN are given by
\eqna{(\psi,\phi)+e&\to(\psi,\phi)+e\\
(\psi,\phi)+N&\to(\psi,\phi)+N.
}

To undertake a full treatment of the relevant physics necessitates the implementation of dark photons and dark sectors in SN simulation codes, an endeavor which is beyond the scope of this work.  In the following we instead follow~\cite{Dreiner:2003wh} and rely on two analytic criteria.  The first demands that the integrated emitted energy by the SN through the dark sector channel $E_\text{D}$ is less than about a tenth of the emitted energy through neutrinos, \textit{i.e.}
\eqn{E_\text{D}<E_\text{D}^\text{max}=10^{52}\ \text{erg} \simeq \frac{1}{10} E_\nu \, .
}[Ebound]
The second, the so-called Raffelt criterion, requires that the emissivity in dark sector particles $\dot{\mathscr{E}}_\text{D}$ does not alter the neutrino signal observably, \textit{i.e.} 
\eqn{\dot{\mathscr{E}}_\text{D}<\dot{\mathscr{E}}_\text{D}^\text{max}=10^{19}\ \text{erg}\cdot\text{g$^{-1}$}\cdot\text{s$^{-1}$}.
}[Emissbound]
This comes from the following reasoning~\cite{Raffelt:1996wa}: at about 1 s after the core bounce the neutrino luminosity is $L_\nu \sim 3\times 10^{52}$ erg$\,\cdot\,$s$^{-1}$. The mass of the object is $M\simeq 3\times 10^{33}$ g. Thus, in order to affect the total cooling time scale, a novel cooling agent would have to compete with the energy-loss rate $L_\nu / M \simeq 10^{19}$ erg$\,\cdot\,$g$^{-1}\,\cdot\,$s$^{-1}$. For the case of an additional energy loss via extra dimensions, the Raffelt criterion was demonstrated to be reliable by a comparison
with results from explicit SN simulations followed by a statistical analysis in Ref.~\cite{Hanhart:2001fx}.

The integrated emitted energy criterion \Ebound is usually more reliable than the Raffelt criterion \Emissbound.  However the latter is easier to implement since it does not require as many integrals to be performed.  In the following we show that both criteria lead to approximately the same constraints, thus increasing our confidence in the simpler Raffelt criterion.

This paper does not deal with the production of dark sector particles from nucleon--nucleon collisions.  
The results of this more involved study that will be based on the formalism of Refs.~\cite{Hanhart:2000ae,Hanhart:2000er, Dreiner:2003wh} will be discussed elsewhere.  For related work involving only a dark photon, but no dark sector, see Refs.~\cite{An:2013yfc,Redondo:2013lna,An:2013lea}.


\section{Dark Forces and Dark Sectors}\label{Dark}

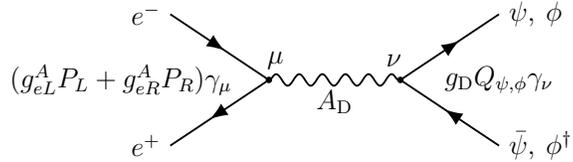
\begin{figure}[!t]
\centering
\resizebox{8cm}{!}{
\begin{tikzpicture}[thick]
\begin{scope}
\draw[-<-](1.5,1)--(0,2) node[left]{$e^-$};
\draw[->-](1.5,1)--(0,0) node[left]{$e^+$};
\draw[->-](3.5,1)--(5,2) node[right]{$\psi$, $\phi$};
\draw[-<-](3.5,1)--(5,0) node[right]{$\bar{\psi}$, $\phi^\dagger$};
\node at (-0.75,1) {$(g^A_{eL}P_L+g^A_{eR}P_R)\gamma_\mu$}; 
\draw[solid,decorate,decoration=snake](1.5,1)--(3.5,1)node[midway,below]{$A_\text{D}$};
\node at (5,1){$g_\text{D} Q_{\psi,\phi} \gamma_\nu$};
\node at (1.6, 1.3) {$\mu$};
\node at (3.4, 1.3) {$\nu$};
\filldraw[black](1.5,1) circle(1pt);
\filldraw[black](3.5,1) circle(1pt);
\end{scope}
\end{tikzpicture}
}
\caption{\textit{Light dark sector particle production mechanism in SN.}  The Feynman diagram represents the relevant production mechanism for electron-positron pair annihilation into light dark sector particles through a dark photon exchange. The couplings $g^A_{eL}$ and $g^A_{eR}$, whose definition is given in the Appendix, are proportional to the small mixing parameter $\varepsilon_Y$. $P_{L,R} = \tfrac{1}{2}(1 \mp \gamma_5)$ are the left and right projectors. Note that contrary to WDs, the dark photon cannot be integrated out.}
\label{FigDark}
\end{figure}

In this section we briefly review the formalism for theories with dark forces and dark sectors. The reader can find more details in Appendix~\ref{app:kinmix}. We consider models that include a spontaneously broken $U(1)_{\rm D}$ gauge group, with the corresponding massive dark photon $A_\text{D}^\mu$, and a dark sector $\mathscr{L}_\text{D}$ which communicates with the SM $\mathscr{L}_\text{SM}$ only through kinetic mixing $\mathscr{L}_{\text{SM}\otimes\text{D}}$~\cite{Fayet:1980ad,Fayet:1990wx,Holdom:1985ag}, \textit{i.e.}
\eqn{\mathscr{L}=\mathscr{L}_\text{SM}+\mathscr{L}_\text{D}+\mathscr{L}_{\text{SM}\otimes\text{D}},\hspace{1cm}\text{where}\hspace{1cm}\mathscr{L}
_{\text{SM}\otimes\text{D}}=\frac{\varepsilon_Y}{2}B_{\mu\nu} F_\text{D}^{\mu\nu}.
}[EqLdark]
Here $F_\text{D}^{\mu\nu}\equiv \partial^\mu A_\text{D}^\nu-\partial^\nu A_\text{D}^\mu$ and $B_{\mu\nu} \equiv \partial_\mu B_\nu-\partial_\nu B_\mu$, where $B_\mu$ is the hypercharge gauge boson.
The kinetic mixing can be thought of as generated by loops of very heavy particles, charged both under the hypercharge and the dark gauge group, and is naturally small: $\varepsilon_Y\sim10^{-4}-10^{-3}$. Below the electroweak scale one can define the mixing to be between the SM photon and the dark photon, with the corresponding parameter $\varepsilon=\varepsilon_Y\cos\theta_W$.  Here $\theta_W$ is the weak mixing angle. In a basis where the gauge bosons have canonically-normalized kinetic terms, the kinetic mixing disappears and is replaced by interactions between the electromagnetically-charged SM fields and the dark photon
\eqn{\mathscr{L}
_{\text{SM}\otimes\text{D}} = -A_\mu^{\rm D} (g^A_{\rm SM,L} J^\mu_{\rm SM,L} + g^A_{\rm SM,R} J^\mu_{\rm SM,R}) \, ,
}[SMcurrent]
where the subscripts L and R indicate currents of left-handed and right-handed SM fields, and the couplings are written explicitly in Appendix~\ref{app:kinmix}.
  In other words the SM fields become millicharged under the {\em dark} gauge group~\cite{Cassel:2009pu,Hook:2010tw}. If the dark sector contains particles charged under $U(1)_{\rm D}$ with masses less than about 100 MeV, they can be produced via the process depicted in Fig.~\ref{FigDark} in a SN and contribute to its cooling, provided they escape. 
  
Eq.~\eqref{SMcurrent} could be rewritten as $-A_\mu^{\rm D}  [(g^A_{\rm SM,R} + g^A_{\rm SM,L})J^\mu_{\rm SM,vec} + (g^A_{\rm SM,R} - g^A_{\rm SM,L})J^\mu_{\rm SM,ax}]$, where $J^\mu_{\rm SM,vec}$ is a vector current and $J^\mu_{\rm SM,ax}$ an axial current. One can check from the explicit expressions in the appendix that the axial coupling $(g^A_{\rm SM,R} - g^A_{\rm SM,L})$ is suppressed by a factor of $\frac{m_{A_{\rm D}}^2}{m_Z^2}$ compared to the vector coupling. Thus, for a dark photon much lighter than the $Z$ boson, as is the case of interest in SN, one can safely neglect the axial coupling.

As already mentioned, in this paper we consider only dark sector particle production mechanisms and scattering processes with the electrons and positrons which are present in a SN.  A follow-up work will discuss the inclusion of nucleons.


\section{Electron-Positron Annihilation to Dark Sector Particles}\label{ep}

In this section we closely follow the analysis of Ref.~\cite{Dreiner:2003wh}. We concentrate on the process with dark fermions in the final states, $e^+(p_1) + e^-(p_2) \to \bar\psi(p_3) + \psi(p_4)$. The one with dark bosons, $\phi$, yields numerically similar results. 

\subsection{Emissivity}\label{Emissivity}

The energy emitted per unit time and unit volume is the emissivity
\eqn{\dot{\mathcal E} (m_{A_\text{D}}, \varepsilon, T_c,\eta) \equiv \frac{d{\mathcal E}}{dt} = \int \frac{d^3p_1 d^3p_2}{(2\pi)^6} f_1 f_2 (E_1+E_2) \vert \mathbf{\Delta v} \vert \sigma(e^+ + e^- \to \bar\psi + \psi),}[EqEmiss]
where $E_1 + E_2$ is the energy of the electron positron pair. The Fermi-Dirac distributions are
\eqn{f_i = \frac{1}{e^{(E_i\pm \mu_i)/T_c}+1},
}[FD]
where $\mu_i$ is the chemical potential and $T_c$ is the temperature in the supernova. We define $\eta \equiv \mu / T_c$ as the degeneracy parameter for the electrons. We have neglected the Pauli blocking of the final state fermions. $\vert \mathbf{\Delta v} \vert$ is the absolute value of the relative M\o ller velocity
\eqn{v_{\rm M\o l} = \sqrt{(\mathbf{v}_1 - \mathbf{v}_2)^2 - (\mathbf{v}_1\times \mathbf{v}_2)^2} \xrightarrow{v_i\to 1} (1-\cos \theta),}[moeller]
where $\mathbf{v}_i$ are the velocities of the incoming electron and positron and $\theta$ is the angle between them in SN frame. The cross section is easily computed by applying the Feynman rules shown in Fig.~\ref{FigDark}:
\begin{equation}\label{fullcross}
\begin{split}
& \sigma(e^+ + e^- \to \bar\psi + \psi) =  \\
& \frac{g_\text{D}^2 Q_\psi^2}{6\pi s [(s-m_{A_\text{D}}^2)^2+m_{A_\text{D}}^2 \Gamma_{\rm tot}^2]} \sqrt{\frac{s-4m_\psi^2}{s-4m_e^2}}(s+2m_\psi^2) [(g^{A2}_{eL} + g^{A2}_{eR})(s-m_e^2) + 6 g^A_{eL} g^A_{eR} m_e^2],
\end{split}
\end{equation}
where $s=(p_1 + p_2)^2$ is the center of mass energy squared, $g_\text{D}$ is the $U(1)_\text{D}$ coupling constant, such that $\alpha_\text{D} \equiv \frac{g_\text{D}^2}{4\pi}$, $Q_\psi$ is the charge of $\psi$ under $U(1)_\text{D}$, which we take to be 1 in the following calculations, $m_{A_\text{D}}$ is the dark photon mass, the couplings $g^A_{eL}$ and $g^A_{eR}$ are defined in Eqs~\eqref{geL} and \eqref{geR}, and $\Gamma_{\rm tot}$ is the total decay width of the dark photon.

The main contribution to the integral \eqref{EqEmiss} occurs when the dark photon is on shell. It is instructive to re-derive the cross section for the on-shell case, which leads to a simpler result. The cross section factorizes 
\eqn{ \sigma(e^+ + e^- \to \bar\psi + \psi) = \sigma(e^+ + e^- \to A_\text{D}) \times  {\rm Br}_{A_\text{D} \to \psi\bar\psi}.
}[factorize]
When the condition $m_{A_\text{D}} > 2 m_\psi$ is satisfied, the dark photon decays into $\psi + \bar\psi$ with an almost 100\% branching ratio, in which case ${\rm Br}_{A_\text{D} \to \psi\bar\psi} \equiv \frac{\Gamma_{A_\text{D} \to \psi\bar\psi} }{\Gamma_{\rm tot}} \simeq 1$. This is because the remaining decay channels are into SM particles and are suppressed by the small mixing parameter $\varepsilon^2$. Then we have to compute 
\eqn{ \sigma(e^+ + e^- \to A_D) = \frac{2 \pi m_{A_{\rm D}}}{4 |\vec p_1| s} \delta(s-m_{A_{\rm D}}^2) \vert \mathcal{M}_{e^+e^- \to A_{\rm D}} \vert^2.
}[cross1]
We find
\begin{equation}\label{deltacross}
\begin{split}
& \sigma(e^+ + e^- \to \bar\psi + \psi) =  \\
& \frac{2\pi}{3m_{A_\text{D}}^2 \sqrt{1- \frac{4m_e^2}{m_{A_\text{D}}^2}}}[(g^{A2}_{eL} + g^{A2}_{eR})(m_{A_\text{D}}^2-m_e^2) + 6 g^A_{eL} g^A_{eR} m_e^2] \delta(s-m_{A_\text{D}}^2)\theta(m_{A_\text{D}} - 2m_\psi).
\end{split}
\end{equation}
Here $\theta(m_{A_\text{D}} - 2m_\psi)$ is the Heaviside step function, needed to enforce the kinematical condition for the decay of the dark photon into $\bar\psi + \psi$. In this form the cross section does not depend on $m_\psi$ as long as the condition $m_{A_\text{D}} > 2 m_\psi$ is fulfilled, nor on $\alpha_{\rm D}$ as expected from unitarity. This is the reason why we did not include any dependence on $m_\psi$ and on $\alpha_{\rm D}$ in the emissivity $\dot{\mathcal E}$ of Eq.~(\ref{EqEmiss}). We use Eq.~(\ref{deltacross}) in the calculations of the next section.


\subsection{Integrated Emitted Energy}\label{Energy}

The total energy emitted in the dark fermion channel is 
\eqn{E_\text{D}(m_{A_\text{D}}, \varepsilon) = \int_0^{t_0} dt \int d^3r \dot{\mathcal E} \left(m_{A_\text{D}}, \varepsilon, T_c(\mathbf{r},t),\eta(\mathbf{r},t) \right)
.}[EqEtot1]
We use the temperature and electron degeneracy distributions from Ref.~\cite{Burrows:1986me}, which are given as functions of the enclosed baryon mass. Assuming a constant density, which is an excellent approximation for $t \geq 250$ ms~\cite{Burrows:1986me}, we can convert the $d^3r$ integral to $dM$. We adopt a core radius of $R_c = 13$ km, a mass of $M_{\rm SN} = 1.4 \ M_\odot$ and obtain a density $\rho \simeq 3 \times 10^{14}$ g/cm$^3$.

The distributions are given at various times~\cite{Burrows:1986me} from $t=0$, corresponding to the time when the incoming shock wave stops and bounces outwards again, up to $t = 20$ s. We use $t_0 = 20$ s as the upper limit of our integral, even if we find that most of the energy is emitted during the first second, as was the case in Ref.~\cite{Dreiner:2003wh}. 

\begin{figure}[!t]
\centering
\includegraphics[width = 0.6\textwidth]{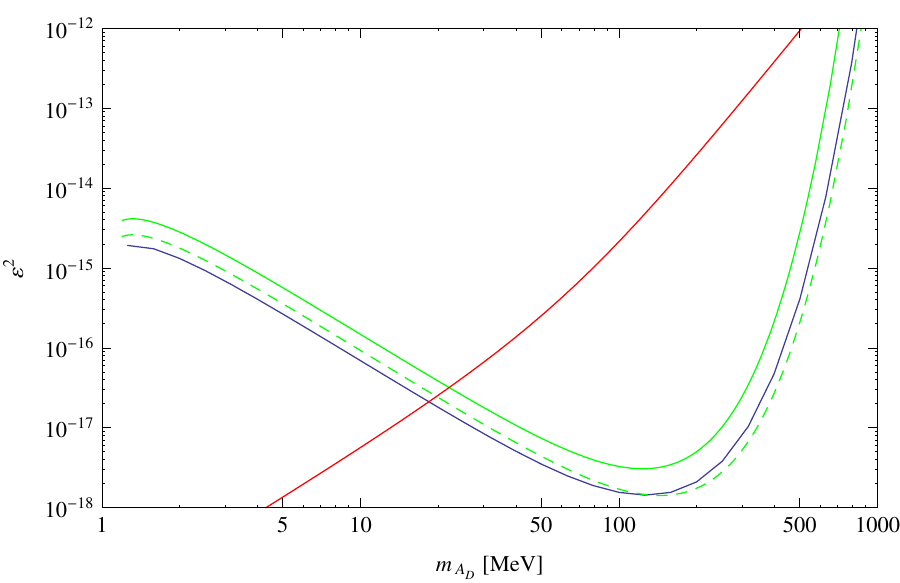}
\caption{ The excluded region is above the blue line, from the cooling constraint of Eq.~(\ref{Ebound}), and below the red line, from the trapping constraint of Eq.~(\ref{depthcrit}). The green lines are obtained with the simpler Raffelt criterion of Eq.~(\ref{Emissbound}), using $T_c = 30$ MeV (plain) and $T_c = 35$ MeV (dashed) for the supernova temperature.  Here $\alpha_\text{D}=10^{-2}$ but note that the cooling constraint is mostly unaffected by the value of the dark fine structure constant.  The dark sector particle mass has to satisfy $m_\psi <\frac{1}{2} m_{A_{\rm D}}$ as explained in the text.}
\label{coolbounds}
\end{figure}

Using the constraint of Eq.~(\ref{Ebound}), $E_\text{D}(m_{A_\text{D}}, \varepsilon)<E_\text{D}^{\rm max}$, we find the lower bound shown in Fig.~\ref{coolbounds} as a blue line.

\subsection{Raffelt criterion}\label{Raffeltcr}

The cooling bound can also be obtained in a computationally simpler way using Eq.~(\ref{Emissbound}), with $ \dot{\mathscr{E}}_\text{D} = \dot{\mathcal{E}}_\text{D} / \rho$. Here the free parameter is the temperature at which $\dot{\mathcal E}$ is to be computed. We use $T_c = 30$ MeV (plain green curve in Fig.~\ref{coolbounds}), as suggested in previous work~\cite{Burrows:1988ah, Hanhart:2000er, Hanhart:2001fx}, and for comparison the higher value $T_c = 35$ MeV (dashed green curve in Fig.~\ref{coolbounds}). Both values result in good agreement with the integrated energy constraint we derived in the previous section.  Thus we are confident that the Raffelt criterion is quite accurate.

Note that the right hand side of Eq.~\eqref{Ebound} could be multiplied by a factor of order 1, which would result in slightly shifting up or down the blue curve in Fig.~\ref{coolbounds}. Due to this arbitrariness one should not take the fact that the dashed green curve (Raffelt criterion) is in better agreement with the blue one as an indication that $T_c=35$ MeV is preferred over $T_c=30$ MeV. The purpose of the plot is simply to show that the simpler Raffelt criterion is a good approximation when compared to the more accurate and elaborate criterion of integrated energy.


\section{Trapping}\label{Trapping}

\subsection{Diffusive Trapping}\label{DiffTrap}

The cooling constraint derived in the previous section applies only if the produced dark particles free stream out of the supernova. To determine whether or not this is the case we consider their mean free path
\eqn{\lambda_\psi = \frac{1}{n_e \sigma_{\psi e \to \psi e}},
}
where $n_e = 8.7\times 10^{43}$ m$^{-3}$~\cite{Burrows:1986me} is the number density of target electrons in the supernova and $\sigma_{\psi e \to \psi e}$ is the cross section for the scattering of dark fermion on electron, which is related via crossing symmetry to the one for the production process $e^+ + e^- \to \bar\psi +\psi$. We use the optical depth criterion~\cite{Burrows:1986me}
\eqn{\int_{r_0}^{R_c} \frac{dr}{\lambda_\psi} \leq \frac{2}{3}
}[depthcrit]
to find if dark particles produced at $r_0$ free stream out of the supernova. Most of the $\psi$'s are produced in the outermost 10\% of the star~\cite{Dreiner:2003wh}, thus we set $r_0 = 0.9 R_c$. The resulting constraint is shown as a red line in Fig.~\ref{coolbounds} for $\alpha_\text{D}=10^{-2}$. In the region above such a line the dark particles are trapped and the simple cooling argument cannot be applied.  In determining the mean free path it is necessary in principle to include the effects of scattering off of nucleons. In the case of protons, the cross section for $\psi p \to \psi p$ is obtained from $\sigma_{\psi e \to \psi e}$ by replacing the electron mass with the proton mass. We have computed this contribution and found that the effect on trapping is negligible compared to $\psi e \to \psi e$. Since the dark photon couples to neutrons even more weakly than to protons we can safely neglect the process $\psi n \to \psi n$.


\subsection{Gravitational Trapping}\label{GRTrap}

Dark sector particles can also be gravitationally trapped in SN.  Again we follow~\cite{Dreiner:2003wh} who showed that relativistic particles almost always escape SN while non-relativistic particles are not gravitationally trapped if their mass is smaller than about $285$ MeV.  Since we are interested in dark sector particles with masses between $0$ and $100$ MeV, the trapping due to gravity is of no consequence.


\section{Results and Conclusions}\label{Results}

\begin{figure}[!t]
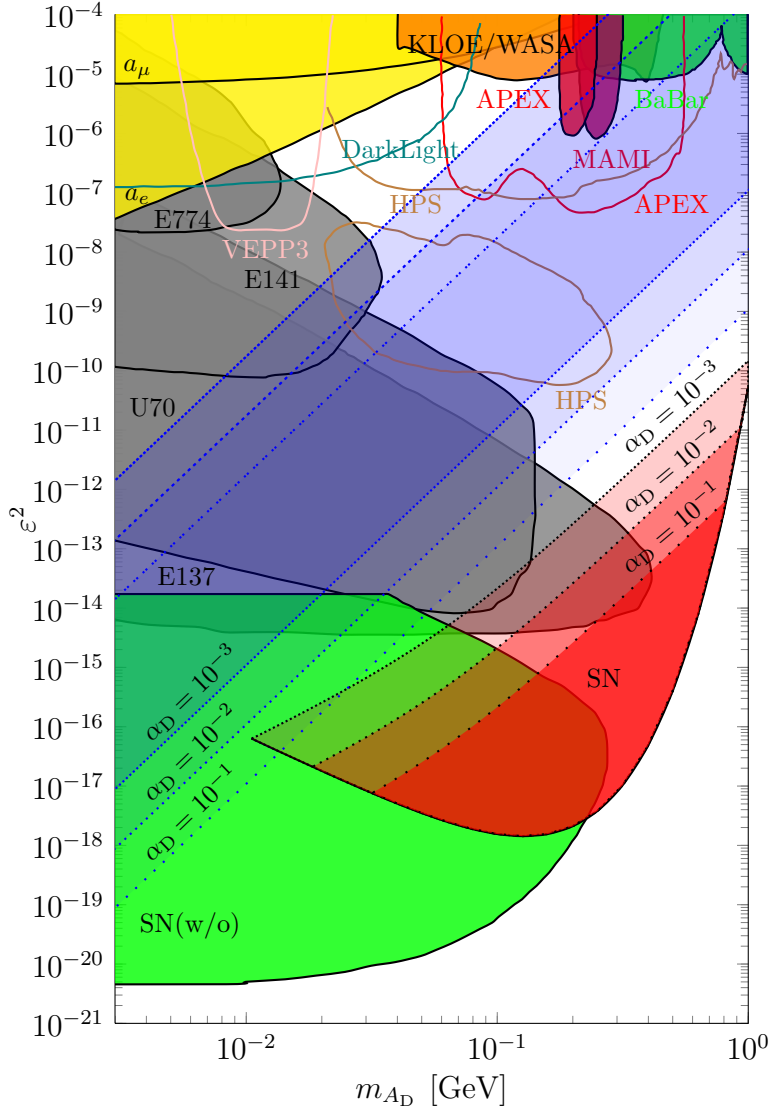

\centering

\caption{\textit{Parameter space exclusion of dark forces with dark sector particles of $\mathscr{O}$(few tens of MeV) from energy losses in SN.}  The red shaded regions are excluded by SN cooling for $\alpha_\text{D}=10^{-1}$ (loosely dotted lines), $\alpha_\text{D}=10^{-2}$ (dotted lines) and $\alpha_\text{D}=10^{-3}$ (densely dotted lines) respectively.  SN trapping constraints forbid the exclusion of regions with larger mixings.  For experiments, which usually assume the dark photon decay is predominantly into the SM, shaded regions correspond to completed direct searches while curves show future reach.  For the electron and muon anomalous magnetic moments, shaded regions are excluded by measurements.  The blue shaded regions are excluded by WD cooling arguments in analogy to SN constraints.  The reader is referred to the text for more details.}
\label{FigExclusion}
\end{figure}

It is interesting to compare the SN constraints on the dark sector parameter space with other constraints, as well as experiments designed to probe such models.  Fig.~\ref{FigExclusion} shows the SN constraints which are valid for dark sector particles with masses less than $1/2 m_{A_{\rm D}}$, thus of the order of $\mathscr{O}(1 - 100\,\text{MeV})$, as well as the WD constraints obtained in~\cite{Dreiner:2013tja} which are valid for masses of a few tens of keV.  The SN constraints coming from models where light dark sector particles do not exist~\cite{Dent:2012mx} are also shown in green and labeled SN(w/o).  Fig.~\ref{FigExclusion} also shows different excluded regions (shaded) of the parameter space as well as regions (curves) that will be explored by future experiments~\cite{Bjorken:2009mm, Essig:2013vha, Izaguirre:2013uxa}.  The experiments include beam dump experiments at SLAC: E137, E141 and E774~\cite{Riordan:1987aw,Bross:1989mp,Andreas:2012mt} as well as the beam dump experiment U70~\cite{Blumlein:2011mv}.  $e^+e^-$ colliding experiments like BaBar~\cite{Aubert:2009au,Bjorken:2009mm} and KLOE~\cite{Archilli:2011zc} are also shown\footnote{The KLOE-2 collaboration presented updated results~\cite{Babusci:2012cr} that represent a slight improvement compared to those in Ref.~\cite{Archilli:2011zc}. Comparable bounds are also found by the WASA-at-COSY Collaboration~\cite{Adlarson:2013eza}.}.  Several fixed-target experiments including APEX~\cite{Abrahamyan:2011gv}, DarkLight~\cite{Freytsis:2009bh}, HPS~\cite{Boyce:2012ym}, MAMI~\cite{Merkel:2011ze} and VEPP-3~\cite{Wojtsekhowski:2012zq} are presented.  Finally, Fig.~\ref{FigExclusion} shows electron ($a_e$) and muon ($a_\mu$) anomalous magnetic moment measurements which constraint the parameter space~\cite{Pospelov:2008zw,Davoudiasl:2012ig,Endo:2012hp}.

From Fig.~\ref{FigExclusion} one can see that the SN constraints of~\cite{Dent:2012mx} and the SN constraints with dark sectors obtained here are in good agreement and complementary.  It is interesting to see that the constraints on dark forces with dark sectors coming from SN is not as strong as could have been expected from the analogous constraint obtained with the help of WDs.  The main reason comes from trapping which is significant in SN. However one should keep in mind that even in the trapping regime there might be constraints when considering the full SN simulation.

Thus, although the constraints on dark forces with dark sectors obtained from WDs are interesting, they suffer from the lightness of the dark sector particles.  On the other hand, the SN constraints allow to probe the dark sector parameter space with masses of the order of a few hundreds of MeV but are rather weak due to trapping.  It is natural to determine the contribution of nucleons in the production of dark sector particles.  Since the nucleons do not affect the trapping considerably, including nucleons might allow to probe smaller values of $\varepsilon^2$.  This possibility will be discussed in a forthcoming publication.


\ack{\vspace{10pt}We thank Adam Burrows and Giuliano Panico for useful discussions.  HD and LU acknowledge the DFG SFB TR 33 ``The Dark Universe'' for support throughout this work.  JFF is supported by the ERC grant BSMOXFORD No.\ 228169. LU would like to thank CETUP* (Center for Theoretical Underground Physics and Related Areas), supported by the US Department of Energy under Grant No. DE-SC0010137 and by the US National Science Foundation under Grant No. PHY-1342611, for its hospitality and partial support during the 2013 Summer Program.}


\appendix

\section{Kinetic mixing} \label{app:kinmix}

\subsection{From gauge to mass eigenstates}

We add to the Standard Model (SM) gauge group an extra $U(1)_\text{D}$, which mixes with the hyper charge $U(1)_Y$. The relevant terms in the Lagrangian are
\beqn
\mathscr{L} &\supset& -\frac{1}{4} (B_{\mu\nu})^2  -\frac{1}{4} (W^3_{\mu\nu})^2
-\frac{1}{4} (F'^{\rm D}_{\mu\nu})^2 - \frac{\sin\varepsilon_Y}{2}B_{\mu\nu}F_{\rm D}'^{\mu\nu} \label{eqlagmix} \\
&+& (D_\mu H)^\dagger (D^\mu H) + \frac{1}{2} m_{A'}^2 A_\text{D}'^2 \label{eqlagmass} \\
&+& i \bar L \gamma^\mu D_\mu L + i \bar e_R \gamma^\mu D_\mu e_R + i \bar Q \gamma^\mu D_\mu Q + i \bar u_R \gamma^\mu D_\mu u_R + i \bar d_R \gamma^\mu D_\mu d_R  \label{eqlagcurrents}  \\
&+&  g_{\rm D} J^{\rm D}_\mu A_{\rm D}'^\mu \label{eqlagdarkcurrents},
\eeqn
where 
\beqn
W^3_{\mu\nu} &=& \partial_\mu W^3_\nu - \partial_\nu W^3_\mu  \label{eqW3munu}\\
B_{\mu\nu} &=& \partial_\mu B_\nu - \partial_\nu B_\mu \\
F'^{\rm D}_{\mu\nu} &=& \partial_\mu A'^{\rm D}_\nu - \partial_\nu A'^{\rm D}_\mu ,
\eeqn
with $B_\mu$ the hypercharge gauge boson, $W^3_\mu$ the third of the $W^a_\mu$'s  $SU(2)_L$ gauge bosons ($a=1,2,3$),  $A'^{\rm D}_\mu$ the $U(1)_{\rm D}$ gauge boson. The prime here denotes the gauge eigenstate. Note that $U(1)_{\rm D}$ is broken and $A'^{\rm D}_\mu$ is massive.  $H$ is the SM Higgs doublet,
\be \label{eqDHiggs}
D_\mu H \supset (\partial_\mu + i g_2 W_\mu^3 \frac{\sigma^3}{2} - i \frac{1}{2} g_1 B_\mu)H.
\ee
In \eqref{eqDHiggs} and \eqref{eqW3munu} we have dropped terms with $W_\mu^1$ and $W_\mu^2$ that are irrelevant to the rest of the discussion here. When the Higgs gets a VEV, $v$, the first term in Eq.~\eqref{eqlagmass} gives us mass terms for $B_\mu$  and $W_\mu^3$.  After rotating to mass eigenstates we will read out the currents from the terms in \eqref{eqlagcurrents} and \eqref{eqlagdarkcurrents}.
In \eqref{eqlagdarkcurrents} the dark current can include fermions and/or bosons
\be
J^{\rm D}_\mu = [Q_\psi \bar\psi \gamma_\mu \psi + Q_\phi i (\phi^\dagger (\partial_\mu \phi) - (\partial_\mu \phi^\dagger)  \phi)].
\ee
In this sector, $g_{\rm D}$ is the gauge coupling constant, with the corresponding $\alpha_{\rm D} \equiv \frac{g_{\rm D}^2}{4 \pi^2}$, $\psi$ and $\phi$ are particles with no SM quantum numbers, but charged under $U(1)_{\rm D}$ with charges $Q_{\psi,\phi}$.

We perform two field-rotations:
\be
\begin{pmatrix}
B_\mu \\
W_\mu^3 \\
A'^{\rm D}_\mu
\end{pmatrix} \quad
\to
\begin{pmatrix}
\tilde B_\mu \\
W_\mu^3 \\
\tilde A_\mu^{\rm D}
\end{pmatrix} \quad
\to
\begin{pmatrix}
A_\mu \\
(A_{\rm NC}^{\rm D})_\mu \\
(\tilde Z_{\rm NC})_\mu 
\end{pmatrix}. 
\ee
With the first one we go from the gauge eigenstates to the fields $\tilde B_\mu$ and $\tilde A_\mu^{\rm D}$ that diagonalize the kinetic terms, with the second one we go to the mass eignestates: $A_\mu$ is the photon (massless), $(\tilde Z_{\rm NC})_\mu$ is {\em almost} the SM $Z$ boson, $(A_{\rm NC}^{\rm D})_\mu$ is what we call the dark photon. The subscript NC stands for Non Canonical, in the sense that these fields do not have canonical kinetic terms, due to the non-unitarity of the first rotation, Eq.~\eqref{eq:firstrotation}. We will have to rescale these fields at the end in order to have them canonically normalized.

Let's begin with the first rotation
\be
\begin{pmatrix}
B_\mu \\
W_\mu^3 \\
A'^{\rm D}_\mu
\end{pmatrix}
=
\begin{pmatrix}
1 & 0 & -\tan\varepsilon_Y \\
0 & 1 & 0 \\
0 & 0 &  \frac{1}{\cos\varepsilon_Y}
\end{pmatrix}
\begin{pmatrix}
\tilde B_\mu \\
W_\mu^3 \\
\tilde A^{\rm D}_\mu
\end{pmatrix}.
\label{eq:firstrotation}
\ee 
This gets rid of the kinetic mixing $- \frac{\sin\varepsilon_Y}{2}B_{\mu\nu}F_{\rm D}'^{\mu\nu}$, the kinetic terms are diagonal now. Eq.~\eqref{eqlagmix} in terms of $\tilde B_\mu$ and $\tilde A^{\rm D}_\mu$ reads
\be
\mathscr{L} \supset -\frac{1}{4} (\tilde B_{\mu\nu})^2  -\frac{1}{4} (W^3_{\mu\nu})^2
-\frac{1}{4} (\tilde F^{\rm D}_{\mu\nu})^2.  \label{eqlagkin}
\ee

Next we substitute $B_\mu = \tilde B_\mu - \tan\varepsilon_Y \tilde A_\mu^{\rm D}$ in Eq.~\eqref{eqDHiggs}. After the Higgs gets a VEV we can read off the following mass matrix  from \eqref{eqlagmass}: 
\be
(\tilde B_\mu \ W_\mu^3 \ \tilde A_\mu^{\rm D})\frac{v^2}{2}
\begin{pmatrix}
\frac{1}{4} g_1^2 & -\frac{1}{4} g_1 g_2 & -\frac{1}{4} g_1\tan\varepsilon_Y \\
-\frac{1}{4} g_1g_2 & \frac{1}{4} g_2^2 & \frac{1}{4}g_1g_2 \tan\varepsilon_Y \\
-\frac{1}{4}g_1^2 \tan\varepsilon_Y & \frac{1}{4}g_1g_2\tan\varepsilon_Y & \frac{1}{4}g_1^2\tan^2 \varepsilon_Y + \frac{m_{A'}^2}{v^2 \cos^2 \varepsilon_Y}
\end{pmatrix}
\begin{pmatrix}
\tilde B^\mu \\
W_3^\mu \\
\tilde A_{\rm D}^{\mu}
\end{pmatrix}
\ee
The mass matrix has determinant zero, as expected due to the residual $U(1)_{\rm EM}$ invariance, so the photon is massless. The other two eignvalues have a complicated form. With the definitions
\beqn
\frac{g_1}{g_2} &\equiv & \frac{s_W}{c_W} \\
m^2_Z &\equiv & \frac{1}{4} v^2 (g_1^2 + g_2^2),
\eeqn
where $s_W$ and $c_W$ are the sine and cosine of the weak mixing angle angle $\theta_W$, they read
\beqn
m_{\tilde Z_{\rm NC}}^2 & = &
\frac{1}{4} \sec ^2\varepsilon_Y \left(2 m_{A'}^2+2 m_Z^2 \left(s_W^2+c_W^2 \cos ^2\varepsilon_Y \right) \right. \\
&+&  \left. \sqrt{2} \left[2 m_{A'}^4+m_Z^2 \left(2 s_W^2 \cos (2 \varepsilon_Y ) \left(m_Z^2 c_W^2-2 m_{A'}^2\right)
       -4 m_{A'}^2 c_W^2 \cos^2\varepsilon_Y  \right. \right. \right.\\
   &-& \left. \left. \left. m_Z^2 (1-2s_W^2)+2 m_Z^2 c_W^4 \cos ^4\varepsilon_Y \right)+m_Z^4\right]^{1/2}\right) \\
m^2_{A^{\rm D}_{\rm NC}} & = &   
\frac{1}{4} \sec ^2\varepsilon_Y  \left(2 m_{A'}^2 +2 m_Z^2 \left(s_W^2+c_W^2 \cos ^2\varepsilon_Y \right) \right. \\
&-&\left. \sqrt{2} \left[2 m_{A'}^4+m_Z^2 \left(2 s_W^2 \cos (2\varepsilon_Y ) \left(m_Z^2 c_W^2-2 m_{A'}^2\right)
      -4 m_{A'}^2 c_W^2 \cos ^2\varepsilon_Y  \right. \right. \right.\\
   &-& \left. \left. \left. m_Z^2 (1-2s_W^2)+2 m_Z^2
   c_W^4 \cos ^4\varepsilon_Y \right)+m_Z^4\right]^{1/2}\right)
\eeqn
Expanding the result for $\varepsilon_Y \ll 1$ we find
\beqn
m_{\tilde Z_{\rm NC}}^2 & = & m_Z^2 \left(1 + \varepsilon_Y^2 \frac{m_Z^2 s_W^2}{m_Z^2 - m_{A'}^2} \right) \\ 
m_{A^{\rm D}_{\rm NC}}^2 & = & m_{A'}^2 \left( 1+ \varepsilon_Y^2 \frac{m_Z^2 c_W^2 - m_{A'}^2}{m_Z^2 - m_{A'}^2} \right).
\eeqn 
From now on all the expressions will be given as expansions up to order $\varepsilon_Y^2$.
The rotation matrix between mass and gauge eigenstates reads
\be
\begin{split}
\begin{pmatrix}
\tilde B_\mu \\
W^3_\mu \\
\tilde A_\mu^{\rm D}
\end{pmatrix}
& =  
\mathscr{R}
\begin{pmatrix}
A_\mu \\
(A^{\rm D}_{\rm NC})_\mu \\
(\tilde Z_{\rm NC})_\mu
\end{pmatrix},  \\
\mathscr{R} & =
\begin{pmatrix}
c_W & \varepsilon_Y \frac{m_Z^2 s_W^2}{m_Z^2 - m_{A'}^2} & m_Z^2 s_W \frac{m_{A'}^4(\varepsilon_Y^2 -1) - \varepsilon_Y^2 m_{A'}^2 m_Z^2 c_W^2 + 2 m_{A'}^2 m_Z^2 + \varepsilon_Y^2 m_Z^4 s_W^2 - m_Z^4}{(m_Z^2 - m_{A'}^2)^3} \\
s_W &   -\varepsilon_Y \frac{m_Z^2 s_W c_W}{m_Z^2 - m_{A'}^2} & - m_Z^2 c_W \frac{m_{A'}^4(\varepsilon_Y^2 -1) - \varepsilon_Y^2 m_{A'}^2 m_Z^2 c_W^2 + 2 m_{A'}^2 m_Z^2 + \varepsilon_Y^2 m_Z^4 s_W^2 - m_Z^4}{(m_Z^2 - m_{A'}^2)^3} \\
0 & 1 - \varepsilon_Y^2 \frac{m_Z^4 s_W^2}{(m_Z^2 - m_{A'}^2)^2} & \varepsilon_Y \frac{m_Z^4 s_W}{(m_Z^2 - m_{A'}^2)^2} 
\end{pmatrix}
\end{split}
\ee
The mass eigenstates, $A_\mu, (A^{\rm D}_{\rm NC})_\mu, (\tilde Z_{\rm NC})_\mu$, have diagonal kinetic terms, but they are not canonically normalized, due to the non-unitarity of the first field transformation~\eqref{eq:firstrotation}. Thus, we perform the following rescalings
\be
\begin{split}
(A^{\rm D}_{\rm NC})_\mu  = & \left( 1 - \frac{\varepsilon_Y^2 m_Z^4}{2(m_Z^2 - m_{A'}^2)^2} + \frac{\varepsilon_Y^2 m_Z^4 (2c_W^2 -1)}{2(m_Z^2 - m_{A'}^2)^2} \right)^{-1/2} A^{\rm D}_\mu \\
(\tilde Z_{\rm NC})_\mu  = & \left( \frac{m_Z^4}{(m_Z^2 - m_{A'}^2)^2} - \frac{\varepsilon_Y^2 m_Z^4 (4m_{A'}^4 - 2 m_{A'}^2 m_Z^2 + m_Z^4)}{2(m_Z^2 - m_{A'}^2)^4} \right. \\
& \left. + \frac{\varepsilon_Y^2 m_Z^6(2c_W^2 -1)(2 m_{A'}^2 + m_Z^2)}{2(m_Z^2 - m_{A'}^2)^4} \right)^{-1/2} \tilde Z_\mu,
\end{split}
\ee
that to order $\varepsilon_Y^2$ do not affect the mass eigenvalues. For the canonical fields, $A^{\rm D}_\mu$ and $\tilde Z_\mu$, we thus have $m_{\tilde Z} = m_{\tilde Z_{\rm NC}}$ and $m_{A_{\rm D}} = m_{A^{\rm D}_{\rm NC}}$. Note that at lowest order the mass eigenvalues correspond to the parameters $m_{A'}$ and $m_Z$.  

\subsection{Couplings of the gauge fields to the currents}

Now we are ready to look at the currents. The covariant derivatives in \eqref{eqlagcurrents} can be written explicitly as 
\beqn
D_\mu L & = & \left(\partial_\mu + i g_2 W_\mu^3 \frac{\sigma^3}{2} - i g_1 \frac{1}{2} B_\mu \right) L \\
D_\mu e_R & = & \left(\partial_\mu  - i g_1  B_\mu \right) e_R \\
D_\mu Q & = & \left(\partial_\mu + i g_2 W_\mu^3 \frac{\sigma^3}{2} + i g_1 \frac{1}{6} B_\mu \right) Q \\
D_\mu u_R & = & \left(\partial_\mu  + i \frac{2}{3} g_1  B_\mu \right) u_R \\
D_\mu d_R & = & \left(\partial_\mu  - i \frac{1}{3} g_1  B_\mu \right) d_R.
\eeqn
We have to express $W_\mu^3$ and $B_\mu$ in terms of the mass eigenstates $A_\mu$, $\tilde Z_\mu$ and $A_\mu^{\rm D}$, using the results derived above. After some algebra, using $g_2 = \frac{e}{s_W}$ and $g_1 = \frac{e}{c_W}$, with $e$ the electric charge, we find the following couplings of the fields to the currents:
\begin{align}
A^{\rm D}_\mu & \left( g_\nu^A \bar \nu_L \gamma^\mu \nu_L + g^A_{eL} \bar e_L \gamma^\mu e_L + g^A_{eR} \bar e_R \gamma^\mu e_R \right. \\
& + g_{uL}^A \bar u_L \gamma^\mu u_L + g_{uR}^A \bar u_R \gamma^\mu u_R + g_{dL}^A \bar d_L \gamma^\mu d_L + g_{dR}^A \bar d_R \gamma^\mu d_R \\
&\left.  + g_{\rm D}^A J^{{\rm D}\mu}\right), \\
 \tilde Z_\mu & \left( g_\nu^Z \bar\nu_L \gamma^\mu \nu_L + g_{eL}^Z \bar e_L \gamma^\mu e_L + g_{eR}^Z \bar e_R \gamma^\mu e_R \right. \\
 & \left. + g_{uL}^Z \bar u_L \gamma^\mu u_L + g_{uR}^Z \bar u_R \gamma^\mu u_R + g_{dL}^Z \bar d_L \gamma^\mu d_L + g_{dR}^Z \bar d_R \gamma^\mu d_R \right. \\
 & \left. + g_{\rm D}^Z J^{{\rm D}\mu}\right),
\end{align}
with
\beqn
g_\nu^A &=& e\varepsilon_Y \frac{m_{A'}^2}{2c_W(m_{A'}^2 - m_Z^2)} \\
g_{eL}^A &=& e\varepsilon_Y \frac{m_{A'}^2 - 2 m_Z^2 c_W^2}{2 (m_{A'}^2 - m_Z^2)}  \label{geL}\\
g_{eR}^A &=& e\varepsilon_Y \frac{1}{c_W}\left( 1 - \frac{m_Z^2 s_W^2}{m_Z^2 - m_{A'}^2} \right) \label{geR}\\
g_{uL}^A &=& - e\varepsilon_Y \frac{m_{A'}^2 - 4 m_Z^2 c_W^2}{6 (m_{A'}^2 - m_Z^2)} \\
g_{uR}^A &=& -\frac{2}{3}e\varepsilon_Y \frac{1}{c_W}\left( 1 - \frac{m_Z^2 s_W^2}{m_Z^2 - m_{A'}^2} \right) \\
g_{dL}^A &=& - e\varepsilon_Y \frac{m_{A'}^2 + 2 m_Z^2 c_W^2}{6 (m_{A'}^2 - m_Z^2)} \\
g_{uR}^A &=& \frac{1}{3}e\varepsilon_Y \frac{1}{c_W}\left( 1 - \frac{m_Z^2 s_W^2}{m_Z^2 - m_{A'}^2} \right) \\
g_{\rm D}^A & = & -\sqrt{4\pi \alpha_{\rm D}} \left( 1 + \varepsilon_Y^2 \frac{m_{A'}^4 -2 m_{A'}^2 m_Z^2 + m_Z^4 c_W^2}{2(m_{A'}^2 - m_Z^2)^2} \right) 
\eeqn
and
\beqn
g_\nu^Z &=& \frac{e}{c_W s_W} \frac{1}{8 (m_{A'}^2 - m_Z^2)^2} \left[ 4 m_{A'}^4 - \varepsilon_Y^2 m_Z^2 (m_Z^2 - 2 m_{A'}^2)\cos(2\theta_W) \right. \nn \\
&{}& \left. \qquad  - 2 m_{A'}^2 m_Z^2 (4+\varepsilon_Y^2) + m_Z^4 (4 + \varepsilon_Y^2)\right] \\
g_{eL}^Z &=&  \frac{-e}{c_W s_W} \frac{1}{16 (m_{A'}^2 - m_Z^2)^2} \left[ \varepsilon_Y^2 m_Z^2 (4m_{A'}^2 + m_Z^2 \cos(4\theta_W) - 3 m_Z^2) \right . \nn \\
& {} & \left. \qquad \qquad + 2 \cos(2\theta_W) (4m_{A'}^4 - 2 m_{A'}^2 m_Z^2 (4 + \varepsilon_Y^2) +m_Z^4 (4 + \varepsilon_Y^2)) \right] \\
g_{eR}^Z & = & \frac{e s_W}{c_W} \left(1 + \frac{\varepsilon_Y^2 m_Z^2 (3 m_Z^2 + m_Z^2 \cos(2 \theta_W) - 4m_{A'}^2)}{4(m_{A'}^2 - m_Z^2)^2} \right)\\
g_{uL}^Z &=& e \frac{1}{12 (m_{A'}^2 - m_Z^2)^2} \left[ 6\cot \theta_W (m_{A'}^2 - m_Z^2)^2 \right. \nn \\
& {} & \left.  - \tan\theta_W (2m_{A'}^4 - 2 m_{A'}^2 m_Z^2 (\varepsilon_Y^2 +2) +2 m_Z^4 \varepsilon_Y^2 \cos(2\theta_W) + m_Z^4 (3\varepsilon_Y^2 +2)) \right]\\
g_{uR}^Z &=& -\frac{1}{6} e \tan\theta_W \left( 4 + \frac{\varepsilon_Y^2 m_Z^2 (3m_Z^2 +m_Z^2 \cos(2\theta_W) - 4 m_{A'}^2)}{(m_{A'}^2 - m_Z^2)^2} \right) \\
g_{dL}^Z &=& -e \frac{1}{12 (m_{A'}^2 - m_Z^2)^2} \left[ 6\cot \theta_W (m_{A'}^2 - m_Z^2)^2 \right. \nn \\
&{}& \left. \qquad - \tan\theta_W (\varepsilon_Y^2 m_Z^4 \cos(2\theta_W) -2(m_{A'}^4 - m_{A'}^2m_Z^2 (\varepsilon_Y^2 +2)+m_Z^4)) \right]\\
g_{dR}^Z &=& \frac{1}{12} e \tan\theta_W \left( 4 + \frac{\varepsilon_Y^2 m_Z^2 (3m_Z^2 +m_Z^2 \cos(2\theta_W) - 4 m_{A'}^2)}{(m_{A'}^2 - m_Z^2)^2} \right) \\
g_{\rm D}^Z & = & \varepsilon_Y \sqrt{4 \pi \alpha_{\rm D}} \frac{m_Z^2 s_W}{m_{A'}^2 - m_Z^2} .
\eeqn

We have not written the couplings of the SM photon, since they are unchanged.

\bibliography{DarkSector-SN}

XXX references not used as of now XXX

\end{document}